\documentclass[
reprint,
superscriptaddress,
 amsmath,amssymb,
 aps,
 prl,
 floatfix,
]{revtex4-2}

\usepackage{graphicx}% Include figure files
\usepackage{epstopdf} %converting to PDF
\usepackage{dcolumn}% Align table columns on decimal point
\usepackage{bm}% bold math
\usepackage{siunitx}
\usepackage{float}
\usepackage{hyperref}
\usepackage{mathtools}

\begin{document}

%\preprint{APS/123-QED}
\title{Tunable Enhancement of Magnetization Dynamics by Crystal Cut at Interface Exchange Coupled $\alpha$-Fe$_2$O$_3$/NiFe Heterostructures}% Force line breaks with \\
%\title{Magnetization dynamics}% Force line breaks with \\
\author{Hassan Al-Hamdo}
\email[E-Mail: ]{h.alhamdo@rptu.de}
\thanks{These authors contributed equally.}
\affiliation{Fachbereich Physik and Landesforschungszentrum OPTIMAS, Rheinland-Pf{\"a}lzische Technische Universit{\"a}t Kaiserslautern-Landau, 67663 Kaiserslautern, Germany}

\author{Tobias Wagner}
\thanks{These authors contributed equally.}
\affiliation{Institute of Physics, Johannes Gutenberg-University Mainz, 55099 Mainz, Germany}

\author{Philipp Schwenke}
\affiliation{Fachbereich Physik and Landesforschungszentrum OPTIMAS, Rheinland-Pf{\"a}lzische Technische Universit{\"a}t Kaiserslautern-Landau, 67663 Kaiserslautern, Germany}

\author{Gutenberg Kendzo}
\affiliation{Fachbereich Physik and Landesforschungszentrum OPTIMAS, Rheinland-Pf{\"a}lzische Technische Universit{\"a}t Kaiserslautern-Landau, 67663 Kaiserslautern, Germany}

\author{Maximilian Dausend}
\affiliation{Fachbereich Physik and Landesforschungszentrum OPTIMAS, Rheinland-Pf{\"a}lzische Technische Universit{\"a}t Kaiserslautern-Landau, 67663 Kaiserslautern, Germany}

\author{Laura Scheuer}
\affiliation{Fachbereich Physik and Landesforschungszentrum OPTIMAS, Rheinland-Pf{\"a}lzische Technische Universit{\"a}t Kaiserslautern-Landau, 67663 Kaiserslautern, Germany}

\author{Misbah Yaqoob}
\affiliation{Fachbereich Physik and Landesforschungszentrum OPTIMAS, Rheinland-Pf{\"a}lzische Technische Universit{\"a}t Kaiserslautern-Landau, 67663 Kaiserslautern, Germany}

\author{Vitaliy I. Vasyuchka}
\affiliation{Fachbereich Physik and Landesforschungszentrum OPTIMAS, Rheinland-Pf{\"a}lzische Technische Universit{\"a}t Kaiserslautern-Landau, 67663 Kaiserslautern, Germany}

\author{Philipp Pirro}
\affiliation{Fachbereich Physik and Landesforschungszentrum OPTIMAS, Rheinland-Pf{\"a}lzische Technische Universit{\"a}t Kaiserslautern-Landau, 67663 Kaiserslautern, Germany}

\author{Olena Gomonay}
\affiliation{Institute of Physics, Johannes Gutenberg-University Mainz, 55099 Mainz, Germany}

\author{Mathias Weiler}
\affiliation{Fachbereich Physik and Landesforschungszentrum OPTIMAS, Rheinland-Pf{\"a}lzische Technische Universit{\"a}t Kaiserslautern-Landau, 67663 Kaiserslautern, Germany}

\begin{abstract}

We investigate spin dynamics in $\alpha$-Fe$_{2}$O$_{3}$/Ni$_{80}$Fe$_{20}$ (Py) heterostructures, uncovering a robust mechanism for designing the ferromagnetic resonance (FMR) frequency through control of crystal orientation, temperature, and applied magnetic field. Employing cryogenic ferromagnetic resonance spectroscopy, we demonstrate that the relative orientation of the N\'eel vector of $\alpha$-Fe$_{2}$O$_{3}$ and the magnetization of the Py layer is highly tunable across the Morin transition temperature ($T_M$). Central to our findings is the pivotal role of crystal orientation in governing the spin dynamics, with the reorientation of the Néel vector serving as a complementary mechanism that further modulates the system's behavior. Our experiments and corroborating theoretical model reveal distinct resonance behavior for different crystal orientations, highlighting the pivotal role of the mutual N\'eel-vector and magnetization geometry in dictating FMR frequencies. In this way, we can tune the FMR frequencies up to a tenfold increase by manipulating the mutual N\'eel-vector / magnetization configuration. Our study comprises all common $\alpha$-Fe$_{2}$O$_{3}$ crystal orientations and develops a unified theoretical description of the ensuing spin dynamics. %These advancements demonstrate the potential for dynamic control of spin interactions in AFM/FM heterostructures, paving the way for the development of advanced spintronic devices with tunable magnetic properties. Our work provides critical insights into the fundamental interactions governing hybrid spin systems and opens new avenues for the design of versatile, temperature-responsive magnetoelectronic applications.

\end{abstract}

\maketitle

Magnetic heterostructures that combine antiferromagnetic (AFM) and ferromagnetic (FM) constituents are widely employed in commercial spintronic devices for sensing and data processing. Continuous research efforts focusing on the physical phenomena underlying the properties of these heterostructures have lead to significant improvements of these devices as well as emerging application scenarios~\cite{li2007exchange, chappert2007emergence, dieny1991giant, nogues1999exchange, skumryev2003beating, nguyen2018magnetoresistive, huang2022spin, lau2016spin, vzelezny2018spin, takano1997interfacial}. 

While the fully compensated magnetic moments of the AFM layer~\cite{neel_proprietes_1936, nagamiya_antiferromagnetism_1955} results in remarkable properties in its own, such as the absence of stray fields and terahertz (THz) spin dynamics~\cite{neel_proprietes_1936, nagamiya_antiferromagnetism_1955, rezende2019introduction}, the key requirement for the functional properties of AFM/FM heterostructures is the coupling of the AFM and FM magnetic order~\cite{meiklejohn1956new, finazzi2004interface, mcmichael1998ferromagnetic, schlenker1968etude, lamirand2013robust, mishra2010dual, takano1997interfacial}. This coupling gives rise to functional magnetic phenomena, including shifts in the hysteresis loop, enhanced coercivity of the FM layer~\cite{meiklejohn1956new,mishra2010dual, lamirand2013robust}, rotational hysteresis under high magnetic fields~\cite{schlenker1968etude}, and rotatable anisotropy~\cite{prosen1961rotatable, lommel1962rotatable}.  The strength of this coupling can be modulated by adjusting the thickness of the FM or AFM layers~\cite{al2023coupling, kohlhepp2004influence, jung2004effect, singh2018thickness, Beiranvand2021, pokhil2002magnetic, beik2017broadband, stoecklein1988ferromagnetic}.  In ferromagnetic resonance experiments, this interfacial coupling also produces effects such as unidirectional anisotropy, increased resonance linewidth, and shifts in the ferromagnetic resonance field and frequency~\cite{scott1985ferromagnetic, mcmichael1998ferromagnetic, al2023coupling, speriosu1987standing, stoecklein1988ferromagnetic}.

Here we experimentally and theoretically demonstrate that the dependence of the effective anisotropy and ferromagnetic resonance frequency on the relative orientation of N\'eel vector and magnetization can be exploited to enable in-situ control of spin dynamics in $\alpha$-Fe$_2$O$_3$/Py heterostructures.

$\alpha$-Fe$_2$O$_3$ is an insulating antiferromagnet (AFM) with a corundum crystal structure~\cite{curry1965neutron, besser1967magnetocrystalline, muench1985morin, chevallier1951proprietes}, characterized by a Morin transition at approximately $T_\mathrm{M} \approx \SI{260}{\kelvin}$~\cite{morin1950magnetic, guilland1951magnetic}. Above $T_\mathrm{M}$, $\alpha$-Fe$_2$O$_3$ behaves as a canted antiferromagnet, exhibiting a small magnetic moment due to the Dzyaloshinskii–Moriya interaction%, where the spins are oriented within the c-plane, perpendicular to the c-axis%
~\cite{schroeer1967morin, muench1985morin}. Below $T_\mathrm{M}$, the spins of $\alpha$-Fe$_2$O$_3$ reorient along the c-axis, leading to the formation of a collinear antiferromagnet. The $T_\text{M}$ can be controlled by doping~\cite{flanders1965magnetic, popov2022influence, krehula2017synthesis}, and $T_\text{M}$ above room temperature has been realized~\cite{shimomura2015morin}.

We study heterostructures based on $\alpha$-Fe$_2$O$_3$ single crystals and \SI{10}{\nano\meter}-thick polycrystalline Py layer prepared by electron beam evaporation (see SI~\cite{SI} for more details). To demonstrate the control of the mutual N\'eel-vector and magnetization geometry, we investigate the ferromagnetic resonance (FMR) behavior of the FM Py layer coupled to both canted and collinear phases of the $\alpha$-Fe$_2$O$_3$ crystals. By varying the AFM crystal orientation and system temperature, we demonstrate that the AFM/FM spin coupling can be exploited to control the FM resonance frequency via reconfiguration of the relative orientation of the N\'eel vector and FM magnetization. Because the N\'eel-vector of $\alpha$-Fe$_2$O$_3$ can be configured in-situ by magnetic fields and temperature, our findings present opportunities for tunable FM dynamics in advanced magnetoelectronic devices.

In our experiments, we utilized AFM/FM heterostructures with a range of $\alpha$-Fe$_2$O$_3$ crystal orientations. In the following, we refer to our heterostructures based on their crystal plane orientations: (11$\overline{2}$0), (10$\overline{1}$0), (1$\overline{1}$02), and (0001) (see SI~\cite{SI}).

The spin dynamics of the uniform FM resonance mode in these samples were studied using Cryogenic Ferromagnetic Resonance Spectroscopy (Cryo-FMR) based on an assembly of a coplanar waveguide connected to a Vector Network Analyzer (VNA) with \SI{43}{\giga\hertz} bandwidth. The investigation was conducted by systematically varying the system temperature from 40 to \SI{400}{\kelvin}.

\begin{figure}\centering
\includegraphics{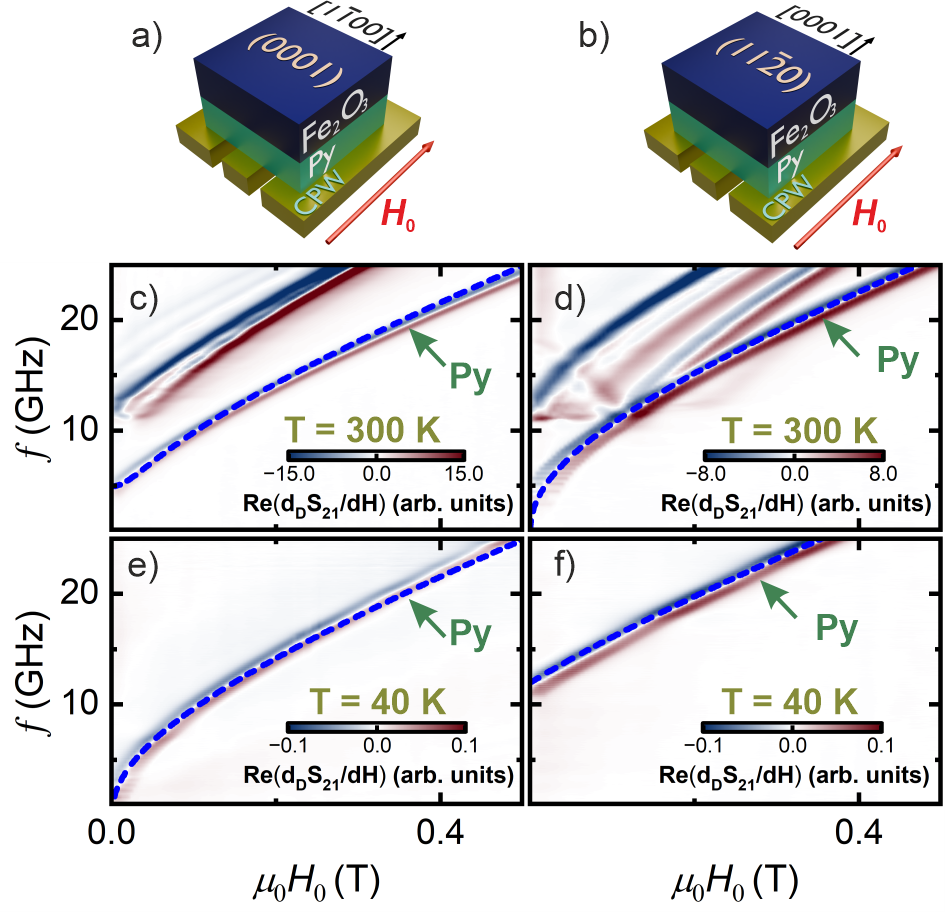} \caption{(a) and (b) Schematic representation of the $\alpha$-Fe$_2$O$_3$/Py heterostructures for the (0001) and (11$\overline{2}$0) samples, respectively. The external magnetic field is applied in-plane for both samples. (c), (e) Real part of the background-corrected VNA signal for the (0001) sample at \SI{300}{\kelvin} and \SI{40}{\kelvin}, respectively.  
(d) and (f) display corresponding measurement results for the (11$\overline{2}$0) sample, where $H_0$ is applied in-plane and parallel to the [0001] axis. The dashed blue lines represent the Py resonance frequencies obtained from our model calculations (see text).\label{fig:figure_1}}
\end{figure}

Figure~\ref{fig:figure_1} presents schematics, experimental data and theoretical modeling for two selected $\alpha$-Fe$_2$O$_3$/Py heterostructures to demonstrate the salient features of N\'eel-vector control of FM spin dynamics.

We first focus on results based on the (0001) heterostructure that is schematically depicted in Fig.~\ref{fig:figure_1}(a). In Fig.~\ref{fig:figure_1}(c), we show the FMR spectra and theoretical model obtained at \SI{300}{\kelvin} - well above the Morin transition of $\alpha$-Fe$_2$O$_3$. Specifically, we show the color-coded $\mathrm{Re}(d\textsubscript{D} S\textsubscript{21}/dH)$ spectra obtained by performing the derivative divide~\cite{maier2018note}  data processing of our raw $S\textsubscript{21}$ VNA-FMR data. Multiple resonance lines above 10 GHz are observed and attributed to the quasi-FMR response of the cuboid-shaped $\alpha$-Fe$_2$O$_3$ crystal~\cite{damon1960magnetostatic}. We observe these modes also in a bare $\alpha$-Fe$_2$O$_3$ crystal without Py layer (see SI~\cite{SI}). In the following, we focus on  the Py resonance because the large volume of the $\alpha$-Fe$_2$O$_3$ crystal prevents the observation of interfacial effects on its spin dynamics.
%Multiple resonance lines above \SI{10}{\giga\hertz} are observed and attributed to the magnetostatic quasi-FMR response of $\alpha$-Fe$_2$O$_3$. These $\alpha$-Fe$_2$O$_3$ resonances are analyzed in more detail in the SI~\cite{SI} and will not be discussed further here in order to focus on the N\'eel-vector control of the Py FMR instead.
At lower frequencies, we observe a further single resonance line in panel (c) that we attribute to the Py FMR response. Noteworthy is, that for $H_0\rightarrow 0$, the FMR frequency is approximately $f=\SI{5}{\giga\hertz}$, while we would expect $f\rightarrow 0$ for an isolated Py film~\cite{zhao2016experimental, klemmer2000ferromagnetic}.

Conversely, Fig.~\ref{fig:figure_1}(e) presents the FMR spectra recorded at \SI{40}{\kelvin}, \textcolor{red}{} where the resonance mode of $\alpha$-Fe$_2$O$_3$ is well outside of the frequency range of our experimental setup due to the transition of $\alpha$-Fe$_2$O$_3$ from canted AFM to collinear AFM. 
Importantly, we observe that the Py FMR frequency is now approaching $f\rightarrow 0$ for $H_0\rightarrow 0$. By comparing to panel (c) we furthermore observe that the Py resonance frequencies at \SI{300}{\kelvin} and \SI{40}{\kelvin} are similar at large $H_0$.

The right column of Fig.~\ref{fig:figure_1} shows the FMR spectra of the (11$\overline{2}$0) heterostructure (see schematic depiction in panel (b)). Again, we recorded the spectra at temperatures above (d) and below (f) $T_\mathrm{M}$, respectively. At \SI{300}{\kelvin}, in contrast to our corresponding observations from the (0001) sample, we observe no enhancement of the Py resonance frequency for $H_0\rightarrow 0$. Rather, for this heterostructure we now observe that the FMR spectra exhibit a clearly enhanced Py resonance frequency  for $T=\SI{40}{\kelvin}$ for all values of $H_0$. The corresponding FMR frequency in panel (f) assumes a value of $f\approx\SI{12}{\giga\hertz}$ for $H_0 = 0$, again in contrast to the corresponding observations from the (0001) sample depicted in panel (e). The linewidth of the Py signal was found to decrease slightly with temperature, with no significant changes observed near or above $T_\text{M}$ (see SI~\cite{SI}).

In summary, the experimental data in Fig.~\ref{fig:figure_1} strongly indicates that the Py magnetization dynamics reacts to the modification of the relative orientation of the magnetic vectors that can be achieved by varying the temperature across the Morin transition and/or changing the interface orientation of $\alpha$-Fe$_2$O$_3$.

In Fig.~\ref{fig:figure_1}, the blue dashed lines represent the result of theoretical modeling which is in agreement with the experimental observations.  In the following, we will discuss our model to give an intuitive understanding of the underlying phenomena.  

We assume a parallel interface coupling between the N\'eel vector ($\mathbf{n}$) of $\alpha$-Fe$_2$O$_3$ and the Py magnetization vector ($\mathbf{m}_F$) with the corresponding interface energy contribution $w_\mathrm{int}=-J_\mathrm{int}\xi\mathbf{n}\cdot\mathbf{m}_F$, where $J_\mathrm{int}>0$ is the coupling strength and $\xi$ is the effective thickness of the interface layer. Here we consider the uniform mode and neglect the inhomogeneous distribution of magnetic order within both Py and $\alpha$-Fe$_2$O$_3$. We also exclude the excitation of the AFM dynamics by focusing on the effective anisotropy of Py induced by the interfacial coupling with $\alpha$-Fe$_2$O$_3$. We also assume that the equilibrium orientation of the N\'eel vector does not depend on the cut and field value, and is either along the [0001] axis (below $T_M$), or in the (0001) plane perpendicular to the external magnetic field (above $T_M$) (see Fig.~\ref{fig:figure_1}). In contrast, the equilibrium orientation of the Py magnetization depends strongly on the orientation of the crystal cut as the stray fields always keep the Py magnetization in the film plane. It is calculated by minimizing the magnetic energy of the Py layer and the energy of the interfacial exchange coupling with the N\'eel vector. The full procedure is detailed in the SI~\cite{SI}.

\begin{figure}
\centering

\includegraphics{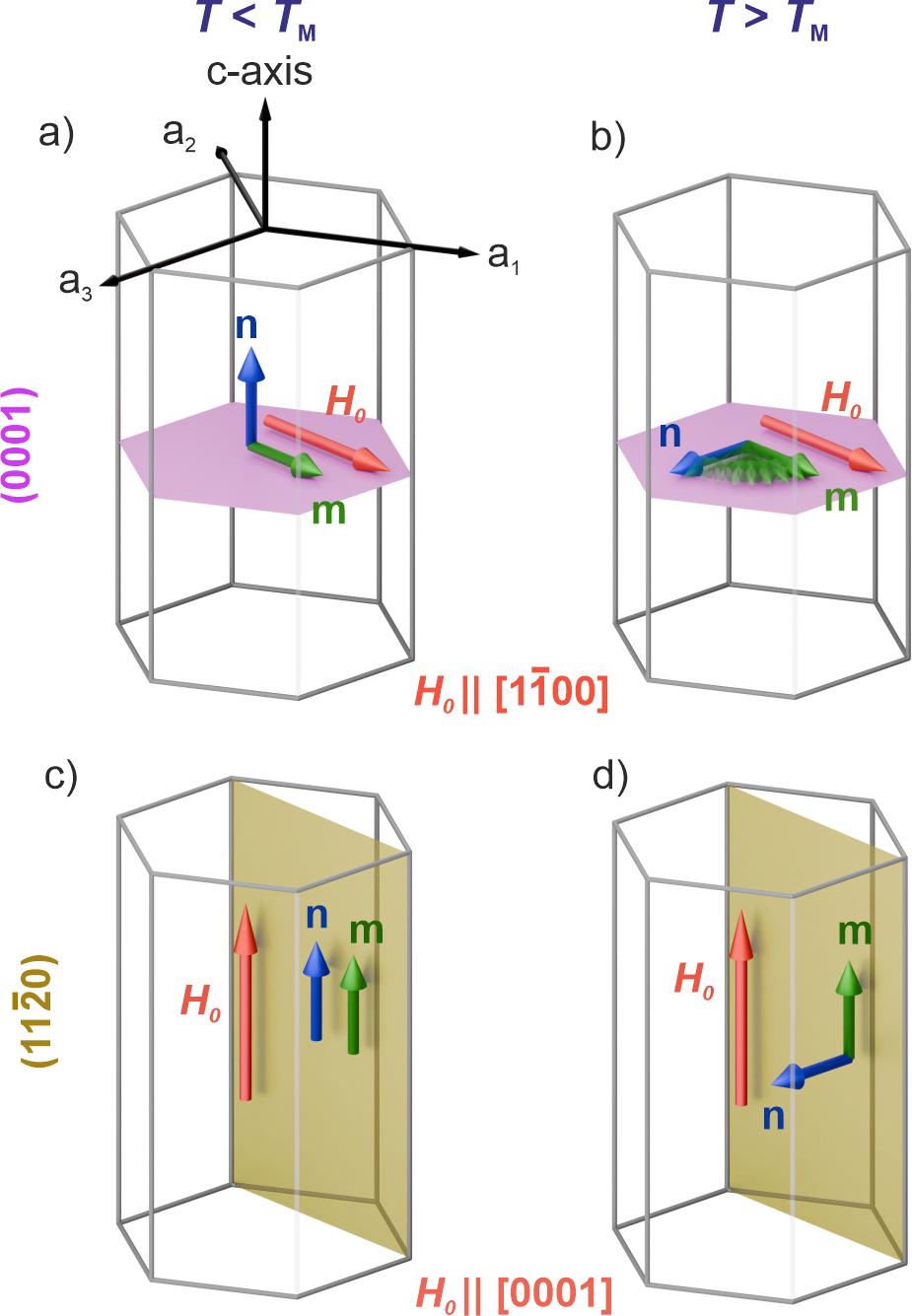}
\caption{Schematic depiction of the orientation of the N\'eel vector of $\alpha$-Fe$_2$O$_3$  ($\mathbf{n}$, blue arrows) and magnetization vector of Py ($\mathbf{m}_\mathrm{F}$, green arrows) below ((a) and (c)) and above ((b) and (d)) the Morin transition for different crystal cuts (shaded areas). Red arrows show the orientation of the external magnetic field. The competition of the Zeemann energy and the interfacial coupling with the N\'eel vector induces a reorientation of the Py magnetization while increasing the magnetic field for $T>T_M$ and (0001) crystal cut as indicated by the transparent green arrows in panel (b).\label{fig:figure2}}
\end{figure}

Within this model, we find that for the (0001) sample above $T_M$ (Fig.\ref{fig:figure2} (b)), the Py magnetization vector is parallel to the N\'eel vector in absence of the magnetic field, $\mathbf{n} \parallel \mathbf{m}_F$. 
The interfacial coupling energy $w_\mathrm{int}$ in this case reaches the minimum value, thus maximizing the coupling.  The external magnetic field $\mathbf{H}_0\|[1\bar{1}00]$ induces a gradual rotation of the Py magnetization away from $\mathbf{n}$, so that the effective coupling is reduced.  

Below $T_M$ (Fig.\ref{fig:figure2} (a)), the N\'eel vector is aligned along the c-axis, perpendicular to both the sample plane and the Py magnetization. The application of an in-plane $H_0$ has no effect on the equilibrium orientation $\mathbf{m}_F$, which always remains orthogonal to $\mathbf{n}$. As a result, in this case $\mathbf{m}_F$ is decoupled from $\mathbf{n}$. 

For the $(11\bar{2}0)$  sample we find a qualitatively different scenario. In contrast to the (0001) sample, $\mathbf{n}$ and $\mathbf{m}_F$ are completely decoupled above $T_M$ (Fig.\ref{fig:figure2} (d)), where $\mathbf{m}_F\perp \mathbf{n}$, and strongly coupled below $T_M$, where $\mathbf{m}_F\| \mathbf{n}$.

We interpret the interface coupling as an effective anisotropy field $H_\mathrm{int}(H_0)$ for the Py magnetization, which is reflected in the field dependence of the FMR frequency:
 \begin{equation}\label{eqn:resonancerfequency}
	f(H_0)=\frac{\mu_0 \gamma }{2\pi}\sqrt{\left[H_\mathrm{int}(H_0)+H_0\right]\left(M_\mathrm{F}+H_0\right)},
\end{equation}
where $\gamma$ is the gyromagnetic ratio of Py, $\mu_0$ is the vacuum permeability and $M_\mathrm{F}$ is the saturation magnetization of Py. The strength of the anisotropy field $H_\mathrm{int}(H_0)= J_\mathrm{int}\xi |\mathbf{n}|\cos\varphi_0(H_0)/d_\mathrm{F}$ depends on the interface exchange coupling and thus decays with the thickness $d_\mathrm{F}$ of the Py layer. The angle $\varphi_0$ between the equilibrium orientation of the Py magnetization and the N\'eel vector depends on the orientation of the cut and on the value of the magnetic field. For the perpendicular orientation $\mathbf{n}\perp \mathbf{m}_F,\mathbf{H}_0$, $\varphi_0=\pi/2$ (Fig.~\ref{fig:figure2}(a) and (d)), $H_\mathrm{int}=0$ and the FMR frequency follows the standard Kittel law resulting in the dashed blue lines in (Fig.~\ref{fig:figure_1}(c) and (e)). In the case of $\mathbf{n}\|\mathbf{m}_F\|\mathbf{H}_0$ (Fig.~\ref{fig:figure2}(c)), $\varphi_0=0$ and $f(H_0)$ dependence shows a gap $\gamma \mu_0\sqrt{H_\mathrm{int}M_\mathrm{F}}$, which explains the enhanced Py resonance frequency reaching up to \SI{12}{\giga\hertz}  (Fig.~\ref{fig:figure_1}(f)). In the intermediate case with $\mathbf{n}$, $\mathbf{m}_F$, and $\mathbf{H_{0}}$ parallel to the interface (Fig.~\ref{fig:figure2}(b)) the angle $\varphi_0$ changes from a non-zero equilibrium value at zero field to $\varphi_0=0$ at saturation. This results in two distinct regions of $f(H_0)$ (see Fig.~\ref{fig:figure_1}(b)) below and above the saturation. For more details of the theoretical model we refer the reader to the SI~\cite{SI}.

\begin{figure}
\centering
\includegraphics{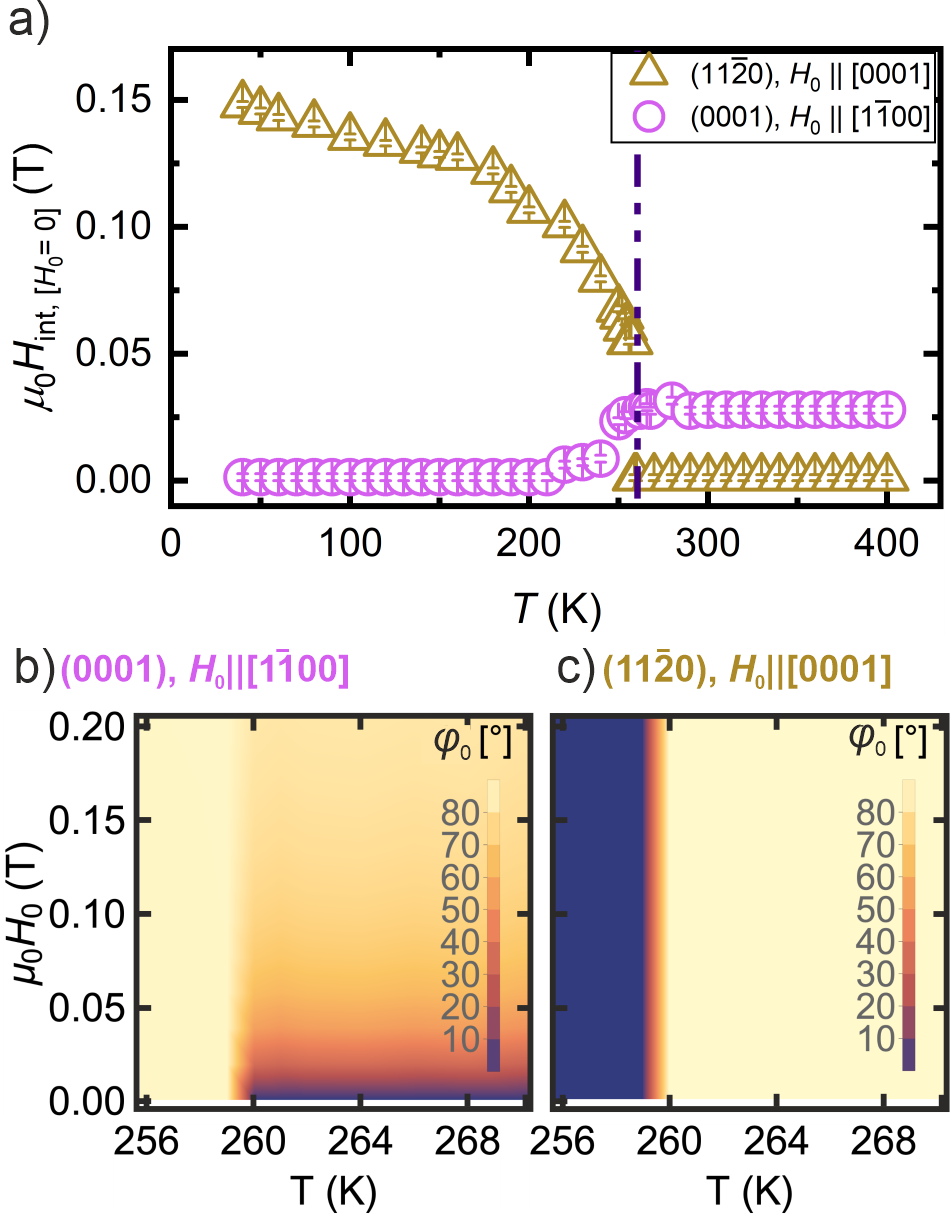}
\caption{a) The effective zero-field interfacial anisotropy, $H_\mathrm{int}(0\,\mathrm{T})$, was determined by fitting the FMR spectra (see Fig.~\ref{fig:figure_1}) with Eq.~\eqref{eqn:resonancerfequency} for the (11$\overline{2}$0) sample with $H_0 || [0001]$(gold triangles) and (0001) sample with $H_0 ||[1\overline{1}00]$ (magenta circles) over a range of temperatures from 40 to \SI{400}{\kelvin}. b) and c) The equilibrium angle $\varphi_0$ between the AFM N\'eel vector and the FM magnetization as a function of the applied static magnetic field and temperature for the (0001)-cut sample (b) and the (11$\overline{2}$0) sample (c). $\varphi_0$ in panels b) and c) was calculated by using the experimentally determined interface exchange coupling strength. They show the complementary behavior of the (0001) and the (11$\overline{2}$0) sample. In each case there are two distinct regions with either low interface exchange coupling (nonzero $\varphi_0$) or strong interface exchange coupling ($\varphi_0=0$). Panel b) additionally showcases the saturation that occurs when the interface exchange coupling competes with the applied magnetic field.}\label{fig:figure_3}
\end{figure}

Figure~\ref{fig:figure_3}(a) shows the effective anisotropy field for vanishing external magnetic field $H_\mathrm{int}(H_0=0)$ that originates from interfacial coupling only, for both samples, measured across a temperature range from \SI{40}{\kelvin} to \SI{400}{\kelvin}. For the (0001) sample, $H_\mathrm{int}$ remains unchanged at temperatures well below $T_M$, indicating no modification in the spin configuration of the layers at the interface in agreement with the schematic of our model shown in Fig.~\ref{fig:figure2}(b). At approximately \SI{220}{\kelvin}, an increase in $H_\mathrm{int}$ is observed, corresponding to the reorientation to parallel alignment of $\mathbf{m}_F$ and $\mathbf{n}$,  and the onset of the effective anisotropy induced by interfacial coupling. 
 
Above this temperature, $H_\mathrm{int}$ increases sharply, reaching the saturation value $\mu_{0} H_\mathrm{int} = \SI{0.03}{\tesla}$ at \SI{270}{\kelvin}.

In contrast, the (11$\overline{2}$0) sample exhibits a maximum $\mu_{0} H_\mathrm{int} =\SI{0.15}{\tesla}$ at the lowest temperature $T=\SI{40}{\kelvin}$  that we set in our experiments (cf. Fig.~\ref{fig:figure2}(d)). When increasing the temperature, near $T_M$, the value of $H_\mathrm{int}$ sharply drops and stays $H_\mathrm{int}\approx0$ for $T>T_M$, indicating decoupling of the Py and $\alpha$-Fe$_2$O$_3$ layers, as shown in Fig.~\ref{fig:figure2}(c).  In the intermediate temperature regime below $T_M$, $H_\mathrm{int}$  smoothly decreases with increasing temperature. We attribute this to hybridization of the Py and $\alpha$-Fe$_2$O$_3$ modes that is not captured in our didactic macrospin model.

We now discuss the impact of non-zero external magnetic fields $H_0>0$ on $H_\mathrm{int}$ and thus the Py FMR frequency obtained from Eq.\eqref{eqn:resonancerfequency}. To this end, Figure~\ref{fig:figure_3}(b) and (c) show the relative orientation of N\'eel vector and magnetization parametrized by $\varphi_0$ as a function of temperature and $H_0$ as obtained from our model. Figure~\ref{fig:figure_3}(b) shows the case of the (0001) sample, where the applied magnetic field and the N\'eel vector are orthogonal (see Figure~\ref{fig:figure2}(a) and (b)). For low temperatures the (0001) sample shows zero $H_\mathrm{int}$ regardless of the strength of the applied field, as the N\'eel vector lies perpendicular to the sample plane. Above the Morin transition the N\'eel vector lies in the sample plane, but perpendicular to the applied field. For low applied fields we observe $\varphi_0=0$ and strong interface exchange coupling $H_\mathrm{int}$. As the applied magnetic field is increased, the FM magnetization rotates away from the AFM N\'eel vector and into alignment with the applied magnetic field, reducing $H_\mathrm{int}$.

Figure~\ref{fig:figure_3}(c) shows the complementary case in the (11$\overline{2}$0) sample. Here, below the Morin temperature the  N\'eel vector is parallel to the applied field (cf. Fig.~\ref{fig:figure2}(d)). Thus, we observe $\varphi_0=0$ and strong interface exchange coupling $H_\mathrm{int}$ with concomitant enhanced Py resonance frequency for all applied field strengths in Fig.~\ref{fig:figure_1}(f). Above the Morin temperature, the N\'eel vector again points out of the sample plane such that $\varphi_0=\pi/2$. This results in vanishing $H_\mathrm{int}$ and unmodified Py resonance frequency.

To confirm our model, a systematic investigation was carried out using further $\alpha$-Fe$_2$O$_3$/Py systems with various $\alpha$-Fe$_2$O$_3$ orientations. The external magnetic field was applied in two distinct in-plane directions (see SI~\cite{SI}). The results demonstrate a systematic variation in the Py FMR frequency with changes in temperature, indicating that both the crystal orientation and temperature significantly influence the hybrid spin dynamics of the system.

\begin{figure}
\centering
\includegraphics{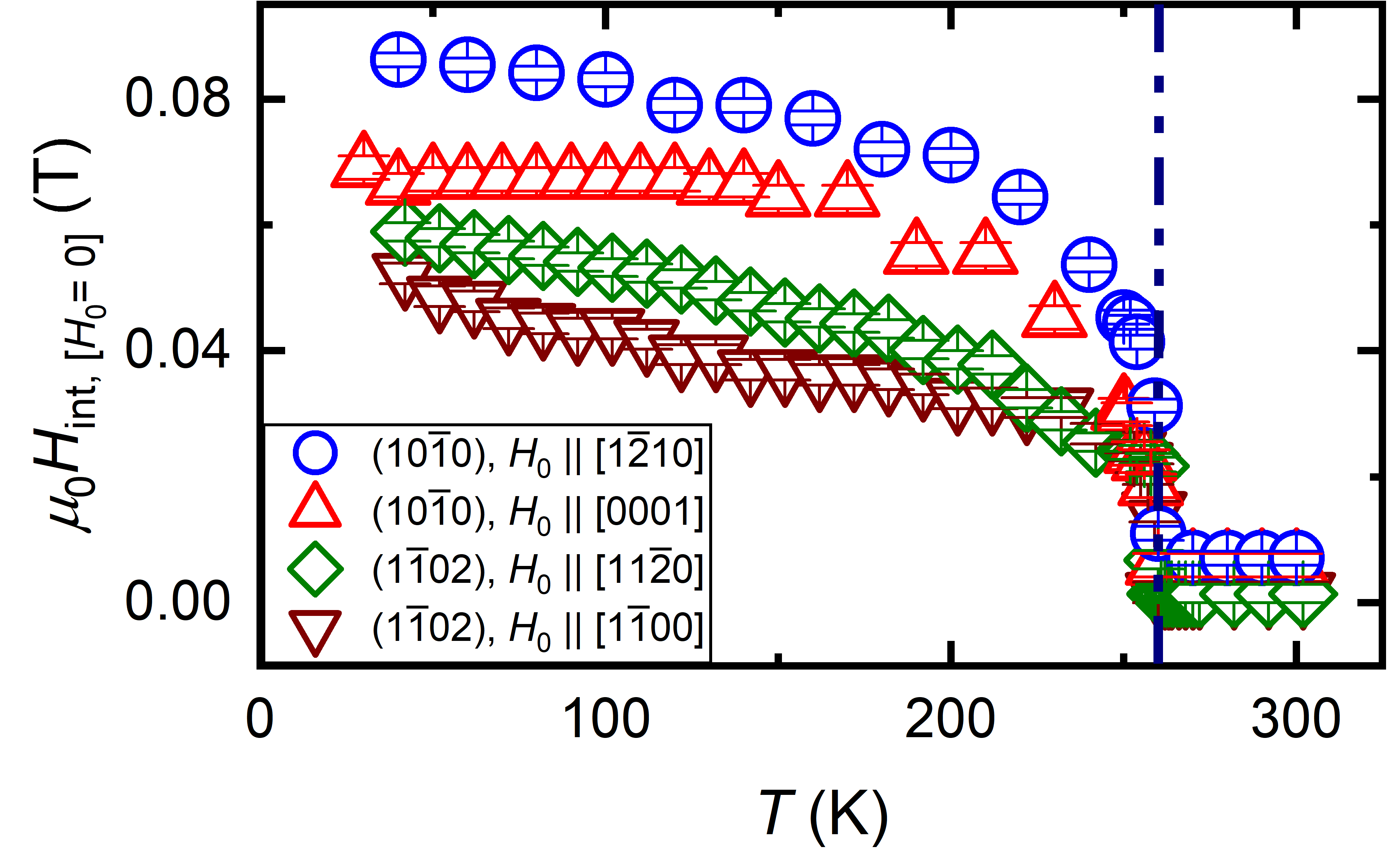}
\caption{The effective anisotropy $H_\mathrm{int}(0)$ for (10$\overline{1}$0) with $H_0 || [1\overline{2}10]$ (blue points), (10$\overline{1}$0) with $H_0 || [0001]$ (red points), (1$\overline{1}$02) with $H_0 || [11\overline{2}0]$ (green points), and (1$\overline{1}$02) with $H_0$ projected along $[1\overline{1}00]$ (brown points). The measurements were conducted over a temperature range of 40 to \SI{300}{\kelvin} (see SI~\cite{SI}).}\label{fig:figure_4}
\end{figure}

Figure \ref{fig:figure_4} presents the effective anisotropy, $ H_\mathrm{int}$, extracted from the fitting of the experimental data for the additional samples analyzed in this study. The samples exhibit behavior similar to that of the \((11\overline{2}0)\) orientation with \( H \parallel [0001] \) (as shown in Fig.~\ref{fig:figure_3}): strong coupling below $T_\mathrm{M}$ with $ H_\mathrm{int}$ reaching maximum value at 40 K, and zero coupling above $T_\mathrm{M}$. Switching between the coupled and decoupled regime is controlled by the temperature induced reorientation of the N\'eel vector.  These findings confirm that the strength of $H_\mathrm{int} $  in $\alpha$-Fe$_2$O$_3$/Py heterostructures is governed by both the crystal structure of $\alpha$-Fe$_2$O$_3$ and temperature.

In conclusion, our investigation of spin dynamics in $\alpha$-Fe$_2$O$_3$/Py heterostructures reveals the critical role of the relative orientation of N\'eel vector and magnetization for spin dynamics in AFM/FM heterostructures. By utilizing cryogenic ferromagnetic resonance spectroscopy, we explored the spin dynamics behavior of these heterostructures as a function of crystal orientation of $\alpha$-Fe$_2$O$_3$ and temperature. Our findings demonstrate that the effective magnetic anisotropy of the ferromagnetic thin film is highly sensitive to both factors, particularly near $T_M$. Our model highlights the importance of the relative alignment between the N\'eel vector in $\alpha$-Fe$_2$O$_3$ and the magnetization vector in Py as key determinants of the spin dynamics.
These results provide valuable insights into the manipulation of spin dynamics in AFM/FM heterostructures, offering potential applications in the design of advanced spintronic devices that leverage the geometric control of the mutual N\'eel-vector / magnetization configuration. %The ability to control the resonance frequency and interfacial coupling  through precise temperature and orientation adjustments presents promising opportunities for future developments in magneto-electronics.\\

\begin{acknowledgments}
We acknowledge financial support by the Deutsche Forschungsgemeinschaft (DFG, German Research Foundation) within the Transregional Collaborative Research Center TRR 173/3–268565370 “Spin+X” (Projects B15, B13 and B11).  O.G. and T.W. also acknowledge support from Deutsche Forschungsgemeinschaft (DFG) SPP 2137, project number 403233384.  M.W. acknowledges support from the European Research Council (ERC) under the European Union’s Horizon Europe research and innovation program (Grant agreement No. 101044526). \\
%H.A. and M.W. contributed to the experimental idea. H.A. performed the experiments, the experimental data post processing and generated the figures containing experimental data. T.W. provided the theoretical model, the data analysis and fitting of the experimental data to the theoretical model and the theory-related figures. H.A., P.S., and G.K. performed the low-temperature measurements. L.S., M.D., and M.Y. contributed to the samples fabrication and characterization. H.A. and T.W. prepared the manuscript. O.G., V.V., P.P. and M.W. provided the scientific leadership of the project and helped with proofreading the manuscript. All authors discussed the results and commented on the paper.
\end{acknowledgments}

\bibliographystyle{naturemag}
\bibliography{Fe2O3-Py}% Produces the bibliography via BibTeX.

\end{document}